\documentclass[prc,aps,floatfix,groupedaddress,amsmath,amssymb,twocolumn]{revtex4}

\usepackage{graphicx}
\usepackage{dcolumn}
\usepackage{bm}
\usepackage{color}

\usepackage[section]{placeins}
\usepackage{graphicx,epsfig}
\usepackage{amsmath}
\usepackage{amsfonts}
\usepackage{braket}
\usepackage{fancyhdr}
\usepackage{hyperref}
\usepackage[utf8]{inputenc}
\usepackage[T1]{fontenc}
\usepackage{amsthm}
\usepackage{hhline}
\usepackage{multirow}
\usepackage[figuresleft]{rotating}
\usepackage{gensymb}
\def\beq{\begin{equation}}
\def\eeq{\end{equation}}
\def\beqa{\begin{eqnarray}}
\def\eeqa{\end{eqnarray}}

\begin{document}

\title{Study of ($p,n$)IAS and ($^3$He,$t$)IAS charge-exchange reactions with the $G$-matrix folding method}

\author{Phan Nhut Huan$^{1,2}$} \email{phannhuthuan@duytan.edu.vn}
\author{Nguyen Le Anh$^3$}\email{anhnl@hcmue.edu.vn}
\author{Bui Minh Loc$^{4,5}$} \email{buiminhloc@tdtu.edu.vn (Corresponding 
author)}
\author{Isaac Vida\~{n}a$^6$} \email{isaac.vidana@ct.infn.it}

\affiliation{
{$^1$} Institute of Fundamental and Applied Sciences, Duy Tan University, Ho 
Chi Minh City, Vietnam \\
{$^2$} Faculty of Natural Sciences, Duy Tan University, Da Nang City, Vietnam \\
{$^3$} Department of Physics, Ho Chi Minh City University of Education, 
Vietnam\\
{$^4$} Division of Nuclear Physics, Advanced Institute of Materials Science, 
Ton Duc Thang University, Ho Chi Minh City, Vietnam \\
{$^5$} Faculty of Applied Sciences, Ton Duc Thang University, Ho Chi Minh City, 
Vietnam \\
{$^6$} INFN Sezione di Catania, Dipartimento di Fisica ``Ettore Majorana'', Universit\`a di Catania,  
Via Santa Sofia 64, I-95123 Catania, Italy 
}

\date{\today}


\begin{abstract}

Differential cross sections of ($p,n$) and ($^3$He,$t$) charge-exchange reactions leading to the excitation of the isobaric analog state (IAS) of the target nucleus are calculated with the distorted wave Born approximation. The $G$-matrix double-folding method is employed to determine the nucleus-nucleus optical potential within the framework of the Lane model.  $G$-matrices are obtained from a Brueckner--Hartree--Fock calculation using the Argonne Av18 nucleon-nucleon potential. Target densities have been taken from Skyrme--Hartree--Fock calculations which predict values for the neutron skin thickness of heavy nuclei compatible with current existing data. Calculations are compared with experimental data of the reactions  ($p,n$)IAS on $^{14}$C at $E_{lab}=135$ MeV and $^{48}$Ca at $E_{lab}=134$ MeV and $E_{lab}=160$ MeV, and ($^3$He,$t$)IAS on  $^{58}$Ni, $^{90}$Zr and $^{208}$Pb at $E_{lab}=420$ MeV.  Experimental results are well described without the necessity of any rescaling of the  strength of the optical potential. A clear improvement in the description of the differential cross sections for the ($^3$He,$t$)IAS reactions on $^{58}$Ni and $^{90}$Zr targets is found when the neutron excess density is used to determine the transition densities.
Our results show that the density and isospin dependences of the $G$-matrices play a non-negligible role in the description of the experimental data.

\end{abstract}

\pacs{}

\maketitle


\section{Introduction}

The charge-exchange reactions ($p,n$) and ($^3$He,$t$) leading to  the excitation of the isobaric analog state (IAS) of the target nucleus (hereafter referred to as CXIAS) are very powerful tools for probing the isospin dependence of the nucleon-nucleus and nucleus-nucleus optical potentials (OP) \cite{Drisko62,Satchler64, Satchler69} and for constraining, in addition,  bulk properties of nuclear matter, particularly, the symmetry energy \cite{Nunes20}. In these reactions, the IAS of the nucleus $^A(Z+1)$  keeps the same structure as the ground state of the target $^AZ$ except for the fact that one of the neutrons has been replaced by a proton, being the excitation energy of the IAS approximately equal to the Coulomb energy of the added proton. The similar structure of the initial and final states makes the  ($p,n$)  and ($^3$He,$t$) reactions very much like ``elastic'' scattering in which the isospin of the incoming proton or $^3$He is flipped \cite{Drisko62,Satchler64, Satchler69}. The isospin flip is driven by the isovector term of the nuclear OP, usually known as Lane potential \cite{Lane62} (see Eq.\ (\ref{Lanepot}) below), which is proportional to the difference of the neutron and proton OPs, and whose knowledge is of fundamental interest for studying nuclear phenomena in which neutrons and protons participate in a different way. Accurate measurements and analysis of the cross sections of CXIAS reactions are very important since the Lane potential together with the folding method \cite{Satchlerbook} provides a direct link between these reactions and the isospin dependence of the in-medium nucleon-nucleon (NN) interaction \cite{Doering75} which is crucial for the understanding of the nuclear matter equation of state and different aspects of the physics of core collapse supernova \cite{Oertel17}, neutron stars \cite{Pharos18} and their mergers \cite{Abbott19}.

First theoretical studies of CXIAS reactions were done by  Satchler {\it et al.} \cite{Drisko62,Satchler64} soon after Anderson {\it et al.} \cite{Anderson61, Anderson61b} identified, in 1961, direct ($p,n$) transitions between IAS in medium weight nuclei. Since then many other studies of CXIAS reactions have been carried out and used to extract the isospin dependence of the nuclear OP. In most of these studies phenomenological optical model potentials have been used to construct the incoming and outgoing scattering states as well as the Lane potential that describes the transition between IAS. There exist several ``global'' sets of the nucleon-nucleus OP parameters derived from extensive optical model analysis of nucleon elastic scattering, such as those of Beccheetti and Greenless \cite{Beccheeti69},  the global OP CH89 of Varner {\it et al.} \cite{Varner91},  and the most recent OP parametrization by Koning and Delaroche \cite{Koning03} which covers a wide range of energies (from 1 keV to 200 MeV) and target masses ($24 \leq A \leq 209$). These global potentials, parametrized in the form of empirical Woods--Saxon functions, are built by simultaneously fitting elastic scattering data for a wide range of stable target nuclei and energies, and are very suitable in predicting the nucleon-nucleus OP when data is not available or cannot measured as it is the case of the unstable nuclei in the drip lines. However, although they are able to reproduce the general trends over different mass and energy regions and are currently used to describe reactions on unstable targets, one must be aware that their validity for unstable nuclei has not been probed. Due to  the large neutron excess in the unstable neutron-rich nuclei it is very important to know the isospin dependence of the nucleon-nucleus OP with the maximum possible accuracy before using it in studies of nuclear reactions and astrophysical phenomena. 

Since high-quality ($p,n$) data  for a wide range of target masses and proton energies is scarce, the isospin dependence of the nucleon-nucleus OP has been mainly deduced (see Refs.\ \cite{Beccheeti69,Varner91,Koning03}) from the analysis of elastic neutron and proton scattering with the same target at about the same energy, taken the strength of the Lane potential equal but with opposite signs for ($n,n$) and  ($p,p$) reactions and treating in some simple way the Coulomb correction terms. In few cases the Lane potential have been also deduced from studies of charge-exchange ($p,n$) scattering to IAS within the distorted-Born wave approximation (DWBA) \cite{Carlson75,Jon00}.  These studies are based on phenomenological nucleon-nucleus OP which, however, are not very well constrained and lead to large uncertainties in the determination of the Lane potential \cite{Lowell17}. Therefore, it is necessary to have a reliable microscopic prediction for the Lane potential to reduce the uncertainty associated with the isospin dependence of the nucleon-nucleus OP. The single-folding method \cite{Satchlerbook}, in which an in-medium or effective NN interaction is folded with the density of the target nucleus, has been employed to construct semi-microscopically the Lane potential, giving rise to a successful description of nucleon elastic and charge-exchange ($p,n$)IAS reactions \cite{Bauge98,Bauge00,Khoa05,Khoa07} after the adjustment of few parameters. The density and isospin dependent effective NN interaction used in Refs.\ \cite{Bauge98,Bauge00} was built from earlier nuclear matter calculations by Jeukenne, Lejeune and Mahaux \cite{JLM74,JLM76,JLM77,JLM77b} whereas that of Refs.\ \cite{Khoa05,Khoa07} was based on the so-called CDM3Y interaction \cite{Khoa93,Khoa97}.

Without any free parameter to be adjusted,  $G$-matrix folding calculations \cite{Brieva77,Brieva77b,Rikus84,Yamaguchi86}, performed with realistic NN interactions that reproduce NN cross sections, phase shifts and the properties of the deuteron, constitute a sophisticated way of deriving microscopic OPs. The so-called Melbourne $G$-matrix \cite{Amos02} has been widely applied to describe elastic and inelastic nucleon-nucleus \cite{Minomo10} and nucleus-nucleus \cite{Egashira14,Toyokawa15} scattering. The important role played by the density and isospin dependence of the effective NN interaction in nucleon-nucleus scattering was already emphasized in the work of Cheon and Takayamagi \cite{Cheon92}. $G$-matrix single-folding calculations for the study of ($p,n$)IAS reactions at proton energies between 80 MeV and 800 MeV were performed by Arellano and Love \cite{Arellano07} using six different NN potential models (Paris, Nijmegen I and II, Reid 93, CDBONN and Argonne Av18) in the construction of the $G$-matrix. Recently, microscopic calculations of nucleon-nucleus and nucleus-nucleus scattering have been also performed using modern chiral two- and three-nucleon forces \cite{Minomo14,Toyokawa15b, Toyokawa15c,Minomo16,Toyokawa18}.
  
Unlike the nucleon-nucleus case, the isovector part of the $^3$He-nucleus OP has been less studied. A relative recent global OP for $^3$He and triton \cite{Pang09}, for instance, does not even contain an isovector term, although it accounts fairly well for the elastic scattering data with a slight dependence of the imaginary part of the OP on the neutron-proton asymmetry of the target nucleus. The analysis of measured data of charge-exchange ($^3$He,$t$)IAS reactions is, therefore, fundamental to determine with accuracy the isospin dependence of the $^3$He-nucleus OP.  Measured data of charge-exchange ($^3$He,$t$)IAS reactions have been mainly studied in the DWBA with the charge-exchange form factor obtained by folding the isospin-dependent part of an effective $^3$He-nucleon interaction with the nuclear transition density for the IAS excitation \cite{Vanderwerf89, Janecker91}. Given the success of the single-folding method in the description of the nucleon elastic scattering and charge-exchange ($p,n$)IAS reactions \cite{Bauge98,Bauge00,Khoa05,Khoa07}, the double-folding (FF) method has been applied to the analysis of  $^3$He elastic scattering and charge-exchange ($^3$He,$t$)IAS reactions. The authors of Ref.\ \cite{Khoa14}, for instance, used the FF method together with a couple channel formalism to describe the elastic $^3$He+$^{14}$C and $^3$He+$^{48}$Ca scattering and the charge-exchange reactions $^{14}$C($^3$He,$t$)$^{14}$N$_{\rm IAS}$ and $^{48}$Ca($^3$He,$t$)$^{48}$Sc$_{\rm IAS}$ at two incident energies of the $^3$He projectile, 72 MeV and 82 MeV in the laboratory frame, employing the isospin- and density-dependent CDM3Y6 effective NN interaction \cite{Khoa97}. On the other hand, in Refs.\ \cite{Loc14,Loc17} the folding method and the DWBA were employed to study charge-exchange ($^3$He,$t$)IAS reactions on  $^{58}$Ni, $^{90}$Zr,  and $^{208}$Pb targets at a projectile energy of 420 MeV. In these two works, a non-relativistic density-independent $T$-matrix, obtained from the phenomenological Franey and Love (FL) effective NN interaction \cite{Love81,Franey85}, is employed in the folding procedure to determine the $^3$He-nucleus OP.

In the present work we use the $G$-matrix double-folding method to determine the nucleus-nucleus optical potential with the framework of the Lane model and determine the differential cross sections of ($p,n$)IAS and ($^3$He,$t$)IAS reactions. $G$-matrices are obtained from a Brueckner--Hartree--Fock calculation of nuclear matter using the Argonne Av18 NN potential. Calculations are compared with experimental data of ($p,n$)IAS and ($^3$He,$t$)IAS reactions on different targets. The manuscript is organized in the following way. A brief review of the double-folding method is presented in Sec.\ \ref{sec:formalism}. Results are shown and compared with experimental data in Sec.\ \ref{sec:results}. Finally, a summary and the main conclusions of this work are given in Sec.\ \ref{sec:conclusions}.


\section{Brief review of the double-folding method}
\label{sec:formalism}

The  formalism employed in our study of ($p,n$)IAS and ($^3$He,$t$)IAS reactions is briefly reviewed here. The interested reader 
can find a detailed description of it in Refs.\ \cite{Fesh62, Satchler83}. The central nucleon-nucleus or nucleus-nucleus optical potential 
can be written in general, see Ref.\ \cite{Lane62}, as  
\begin{equation} 
 U(E,\vec R) = U_0(E,\vec R) + 4U_1(E,\vec R) \frac{\hat T_a \cdot  \hat T_A } {aA} \ ,
\label{Lanepot}
\end{equation}
where $E$ is the total energy in the projectile-target center-of-mass frame, $R$ is the relative coordinate between the projectile and the target,  $U_0(E,R)$ and $U_1(E,R)$ are the  isoscalar and the isovector terms of the optical potential, and $\hat T_a$ and $\hat T_A$ are, respectively, the isospin operators of the projectile and the target nucleus, being
$a$  and $A$ their corresponding mass numbers.

 In the DWBA the transition amplitude for the  CXIAS reactions is given by
\begin{equation}
\mathcal{T}^{\rm {DWBA}}(E) = 
 - \frac{2\sqrt{2T_{z_A}}}{aA }
\langle \chi_{\tilde{a} \tilde{A}} |U_1(E,\vec R)| \chi_{aA}\rangle \ ,
\label{T_DWBA}
\end{equation}
with $T_{z_A} = (N-Z)/2$ being third component of the isospin of the the target in the ground state.
The distorted incoming  ($\chi_{aA}$) and outgoing ($\chi_{\tilde{a} \tilde{A} }$) waves  
are  obtained from the solution of the Schr\"{o}dinger equations
\begin{widetext}
\begin{equation}
\left[ K_a + U_0(E,\vec R) - \frac{2 T_{z_A}}{aA} U_1(E,\vec R) + V_C(\vec R) - E_a \right]\chi_{aA}(\vec R) = 0 
\label{distortedwaves1}
\end{equation}
and
\begin{equation}
\left[K_{\tilde{a}} + U_0(E, \vec R) + \frac{2 (T_{z_A} - 1)}{aA} U_1(E, \vec R) + \Delta_C -
E_a \right] \chi_{\tilde{a}\tilde{A}}(\vec R) = 0 \ .
 \label{distortedwaves2}
\end{equation}
\end{widetext}
Here $K_{a(\tilde{a})}$ is the kinetic energy operator, $V_C(\vec R)$ is the Coulomb potential, and $\Delta_C$ is the 
Coulomb displacement energy.  

To determine $U_0(E, \vec R)$ and $U_1(E, \vec R)$ we use the double-folding method  \cite{Toyokawa15,Khoa00}: 
\begin{widetext}
\begin{eqnarray}
 U_0(E,\vec R) &=& \iint  \left [ \rho_a(\vec r_a)v_{0}^{\rm{DR}} (E, \rho,s) \rho_A(\vec r_A) \right. \nonumber \\
 &+& \left.\rho_a(\vec r_a,\vec r_a+\vec s) 
v_{0}^{\rm{EX}} (E, \rho, s) j_0\left(\frac{\vec k(E,\vec R)\vec s}{M_{aA}}\right) \rho_A(\vec r_A,\vec r_A-\vec s  )\right] d^3r_a d^3r_A  
\label{folcalIS}
\end{eqnarray}
and
\begin{eqnarray}
 U_1(E,\vec R) &=& \frac{1}{\beta}\iint  \left [ \Delta \rho_a(\vec r_a)v_{1}^{\rm{DR}} (E, \rho,s) \Delta\rho_A(\vec r_A) \right. \nonumber \\
 &+& \left. \Delta\rho_a(\vec r_a,\vec r_a+\vec s) 
v_{1}^{\rm{EX}} (E, \rho, s) j_0\left(\frac{\vec k(E,\vec R)\vec s}{M_{aA}}\right) \Delta\rho_A(\vec r_A,\vec r_A-\vec s  )\right] d^3r_a d^3r_A  \ ,
\label{folcalIV}
\end{eqnarray}
\end{widetext}
where $\vec s = \vec r_A - \vec r_a + \vec R$, $\rho_i(\vec r, {\vec r}\,')=\rho_i^n(\vec r, {\vec r}\,')+\rho_i^p(\vec r, {\vec r}\,')$ and $\Delta\rho_i(\vec r, {\vec r}\,')=\rho_i^n(\vec r, {\vec r}\,')-\rho_i^p(\vec r, {\vec r}\,')$ are, respectively the one-body isoscalar and isovector density matrix of the $ith$ nucleus, $\rho_i(\vec r)=\rho_i(\vec r, \vec r)$,  $\beta=(N-Z)/A$ is the neutron-proton asymmetry of the target, $M_{aA}=aA/(a+A)$ and $j_0(x)$ is the zero-order spherical Bessel function. 
The relative momentum $k(E,R)$ is determined self-consistently from  
\begin{equation}
 k^2 (E,\vec R) = \frac{2\mu}{\hbar^2} [E - V(E,\vec R)- V_C(\vec R)] \ ,
\end{equation} 
with $\mu$ being the reduced mass of the projectile and the target, and $V(E,\vec R)$ the real part of the optical potential.

The isoscalar and isovector direct ($v_{0}^{DR}, v_{1}^{DR}$) and exchange ($v_{0}^{EX}, v_{1}^{EX}$) terms of the effective NN interaction  are obtained from the following linear combinations of  the four different spin-isospin (ST) channels
\begin{align} 
v_{0}^{\rm{DR}} &= \frac{1}{16} \left[ v^{(00)} + 3 v ^{(10)} + 3 v^{(01)} + 
9v^{(11)}\right] \ , \\
v_{1}^{\rm{DR}} &= \frac{1}{16} \left[-v^{(00)} - 3 v^{(10)} + v^{(01)} + 
3v^{(11)}\right] \ , \\
v_{0}^{\rm{EX}} &= \frac{1}{16} \left[-v^{(00)} + 3 v^{(10)} + 3v^{(01)} - 
9v^{(11)}\right] \ , \\
v_{1}^{\rm{EX}} &= \frac{1}{16} \left[ v^{(00)} - 3 v^{(10)} + v^{(01)} - 
3v^{(11)}\right] \  .
\end{align}

The form of $v^{(ST)}$ in coordinate space is assumed to be a sum of four Yukawa-functions
\begin{equation}
v^{(ST)} (E, \rho, s) = \sum_{i=1}^4 S^{(ST)}_i(E,\rho) \left[\frac{e^{-\mu_i 
s}}{s}\right]  \ ,
\end{equation}
where $S^{(ST)}_i(E, \rho)$, with $\rho$ taken as $\rho=\rho_A(\vec{r}_A - \frac{\vec{s}}{2}$), are complex energy and density dependent strengths, and  $\mu_i$ are the inverse ranges of the effective interaction.  Following the procedure described in detail by Amos {\it et al.} in Ref.\ \cite{Amos02} the strengths (real and imaginary parts) and the inverse ranges are determined numerically by mapping the NN effective interaction with the complex nuclear matter $G$-matrices obtained from a Brueckner--Hartree--Fock calculation using the Argonne Av18 NN potential. We note that in the case of ($p,n$)IAS reactions the projectile density is taken $\rho_a(\vec r_a)=\delta(\vec r_a)$ and, therefore, Eqs.\ (\ref{folcalIS}) and (\ref{folcalIV}) reduce to the ones of the single-folding case.
 
We finish this section by noticing that in this work all the one-body density matrices are localized following the procedure explained in Ref.\  \cite{Brieva77b} (see in particular  Eq.\ (24) of this reference). The nuclear (local) densities of the targets  are  taken from the Skyrme--Hartree--Fock (SHF) calculations  of Ref.\ \cite{Goriely07} whereas the $^3$He and triton densities are taken from the three-body calculation with the Av18 NN interaction of Ref.\ \cite{Nielsen01}.  We would like to mention also that the DWBA calculations were performed using the ECIS06 code \cite{Raynal06} whereas the double-folding ones have been done with the GDF code \cite{Loccode}.


\section{Results and discussions}
\label{sec:results}

\begin{figure}[t!]
\begin{center}
\includegraphics[width=7.0cm,angle=0,clip]{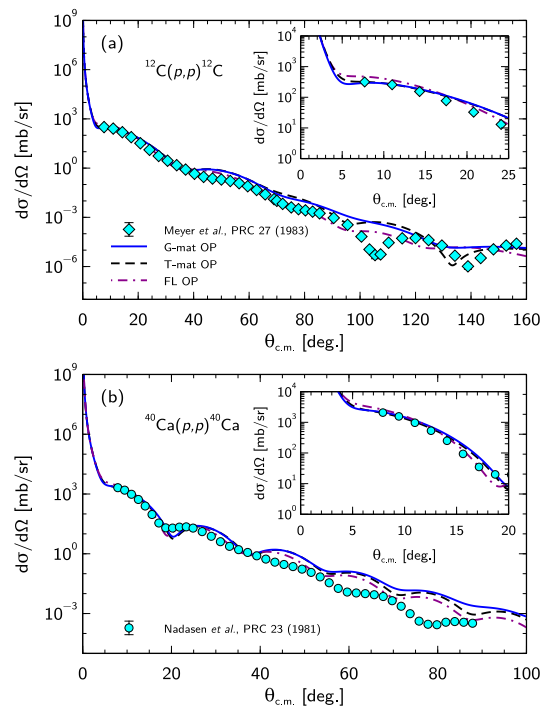}
\caption{Differential cross section for the elastic scattering of a proton on $^{12}$C (panel a) and $^{40}$Ca (panel b) at $E_{lab}=160$ MeV. The two insets show a zoom of the results for low angles.
Results are presented for calculations done with the $G$-mat OP, $T$-mat OP and FL OP, and compared with experimental data from Refs.\ \cite{Meyer83} and \cite{Nadasen81}.}
\label{fig:fig1}
\end{center}
\vskip -0.5cm
\end{figure}

\begin{figure}[t!]
\begin{center}
\includegraphics[width=7.0cm,angle=0,clip]{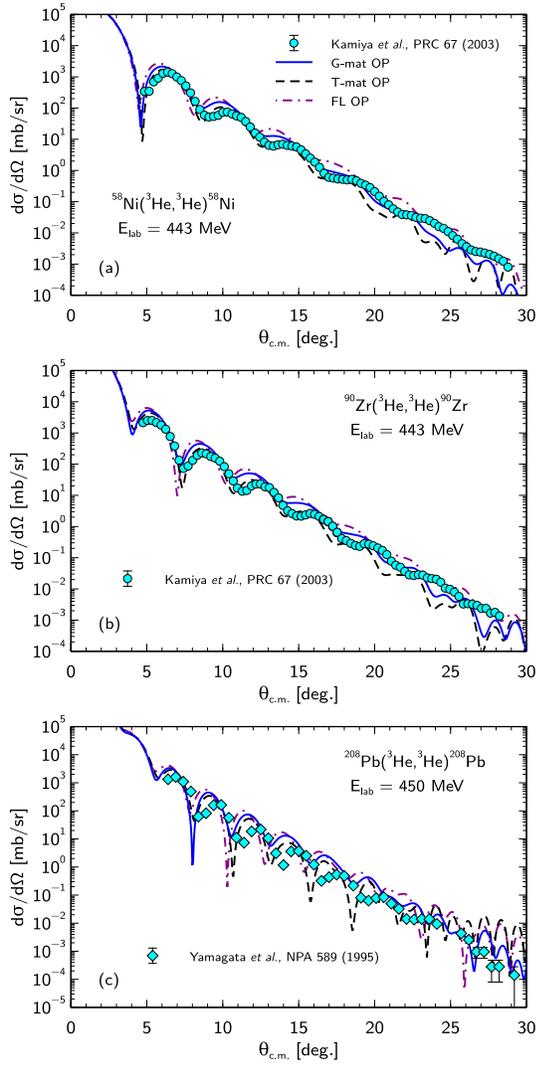}
\caption{Differential cross section for the elastic scattering of $^3$He on $^{58}$Ni (panel a) and $^{90}$Zr (panel b) at $E_{lab}=443$ MeV, and on $^{208}$Pb (panel c) at $E_{lab}=450$ MeV. Results are presented for calculations done with the $G$-mat OP, $T$-mat OP and FL OP, and compared with experimental data from Refs.\ \cite{Kayima01} and \cite{Yamagata95}.}
\label{fig:fig2}
\end{center}
\vskip -0.5cm
\end{figure}


Before analyzing the differential cross section of the charge-exchange ($p,n$)IAS and ($^3$He,$t$)IAS reactions, first, we will briefly comment the results for the elastic channel.  In Fig.\ \ref{fig:fig1} we show the differential cross section for the elastic scattering of a proton on $^{12}$C (panel a) and $^{40}$Ca (panel b) at $E_{lab}=160$ MeV, whereas in Fig.\ \ref{fig:fig2} is presented that of $^3$He on $^{58}$Ni (panel a) and $^{90}$Zr (panel b) at $E_{lab}=443$ MeV, and on $^{208}$Pb (panel c) at $E_{lab}=450$ MeV. 
The results of the calculations, obtained with the $G$-matrix OP with the Av18 NN potential, $T$-matrix OP also derived from the Av18 potential (referred to as $G$-mat, $T$-mat from now on) and FL OP, are compared with experimental data from Refs.\ \cite{Meyer83} and \cite{Nadasen81} in the case of the proton scattering on $^{12} $C and $^{40}$Ca, respectively, and Ref.\ \cite{Kayima01} for the reaction of $^3$He with $^{58}$Ni and $^{90}$Zr and Ref.\ \cite{Yamagata95} for its elastic scattering off $^{208}$Pb.  In the case of the proton elastic scattering off $^{12}$C the density of the carbon target has been taken in this case from Ref.\ \cite{Meyer83} whereas for the other reactions we have used those of Ref.\ \cite{Goriely07}. As it can be seen in both figures, the three calculations describe relatively well the overall magnitude of the experimental cross sections, although there still exists some quantitative disagreement between the calculation and the data. We should note, however, that none of the calculations have been adjusted to reproduce the cross sections. Therefore, it is remarkable the relative good description of the data obtained. The results show that a better description of the data is achieved when using the $G$-mat OP. This indicates that the density dependence of the isoscalar part of the nucleon-nucleus optical potential, which is at the origin of the differences between the cross section obtained using the $G$-mat OP and the $T$-mat OP, plays an important role in the description of the experimental data. Note, however, that  at forward angles the differences in the three calculations, for the five reactions, are very small, meaning this that for those angles the elastic differential cross section is not very sensitive to the density dependence of the OP. As we will see in the following, the effect of the density dependence of the OP becomes more clear in the case of the CXIAS reactions where, in addition, the isospin dependence of the OP plays also a non-negligible role.


\begin{figure}[t!]
\begin{center}
\includegraphics[width=7.0cm,angle=0,clip]{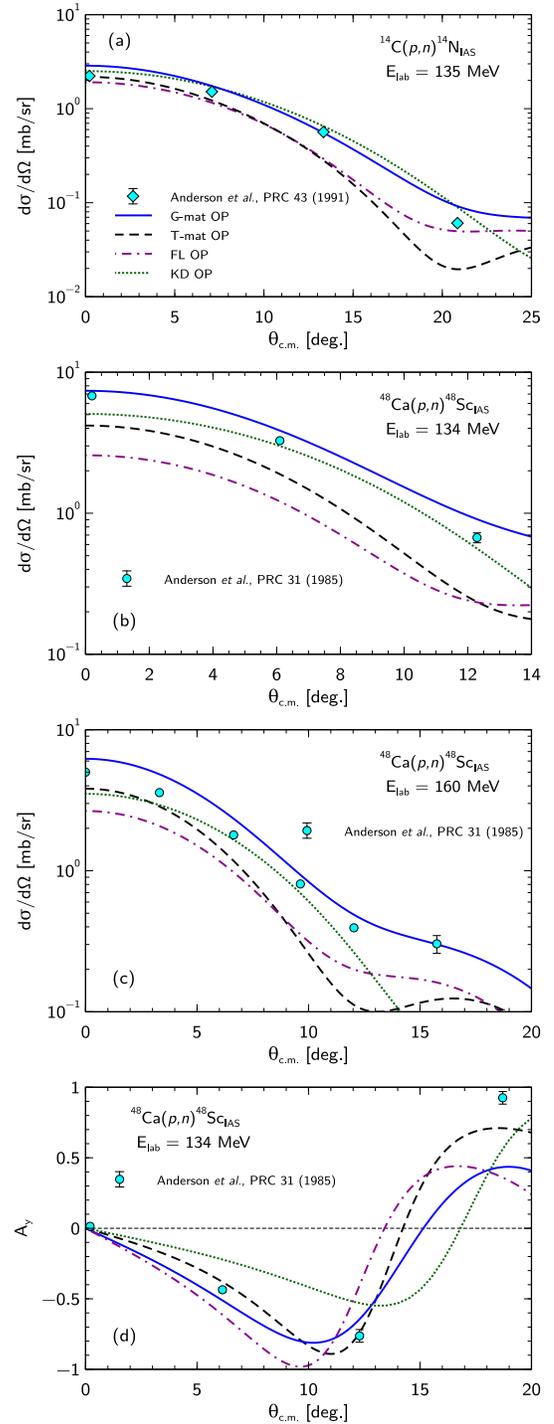}
\caption{Differential cross section of the charge-exchange ($p,n$)IAS reactions on $^{14}$C at $E_{lab}=135$ MeV (panel a) and $^{48}$Ca at $E_{lab}=134$ MeV (panel b) and $E_{lab}=160$ MeV (panel c). Panel d shows the analyzing power for the reaction on  $^{48}$Ca at $E_{lab}=134$ MeV. Results are presented for calculations done with the $G$-mat OP,  $T$-mat OP, FL OP and KD OP, and compared with the experimental data from Refs.\ \cite{Anderson91} and \cite{Anderson85}.
}
\label{fig:fig3}
\end{center}
\vskip -0.5cm
\end{figure}

We show now in Fig.\ \ref{fig:fig3} the differential cross section of the charge-exchange ($p,n$)IAS reactions on $^{14}$C at $E_{lab}=135$ MeV (panel a)  and  $^{48}$Ca at $E_{lab}=134$ MeV (panel b) and $E_{lab}=160$ MeV (panel c). The analyzing power for the reaction on  $^{48}$Ca at $E_{lab}=134$ MeV is also shown in panel d. The results obtained using the proton-nucleus OP derived from the nuclear matter $G$-mat are compared with those in which the OP is determined using a $T$-mat, and the phenomenological OPs derived by Franey and Love \cite{Love81,Franey85}, and Koning and Delarouche (KD) \cite{Koning03}. Experimental data is taken from Refs.\ \cite{Anderson91} for the reaction on the $^{14}$C target and Ref.\ \cite{Anderson85} for that on the $^{48}$Ca one. The general features of the cross sections are qualitatively reproduced by all the calculations, although those performed with the $G$-mat OP gives the best quantitative description of the experimental data. We want to stress that the strength of the $G$-mat and $T$-mat OPs has not been rescaled by any factor in order to obtain a better description of the experimental data. It is, as in the case of the elastic scatteting, remarkable the quality achieved in the description of the experimental data without the necessity of adjusting any free parameter. Note that the results are better described with the $G$-mat OP than with the $T$-mat OP indicating that density dependence of the OP plays a non-negligible role in the determination of the cross sections. In addition, note that the calculations with the $G$-mat OP and $T$-mat OP give also a good qualitative and quantitative description of the  analyzing power for the reaction on $^{48}$Ca at $E_{lab}=134$ MeV. The reason, in part, is simply that the effective NN interaction obtained either from the $G$-matrix or the  $T$-matrix takes completely into account the isospin-dependence of the spin-orbit part of the NN force \cite{Gosset76}.  We observe that the spin-orbit transition potential has been obtained in the same way as the central transition one. To the best of our knowledge this is the first time that the analyzing power is calculated with the folding method for ($p,n$)IAS reactions. Note that the FL OP reproduces qualitatively the shape of the analyzing power but it gives a worse quantitative description. The KD OP has no isospin-dependent spin-orbit NN interaction, because the spin-orbit interaction is taken the same for neutrons and protons in this case. Consequently, the KD OP cannot describe properly the analyzing power, not even qualitatively, simply because the spin-orbit transition potential is missing. Unfortunately, a more comprehensive analysis of the isospin-dependence of the spin-orbit part of the OP is difficult due to the lack of enough experimental data on the analyzing power above 100 MeV.

\begin{figure}[t!]
\begin{center}
\includegraphics[width=7.0cm,angle=0,clip]{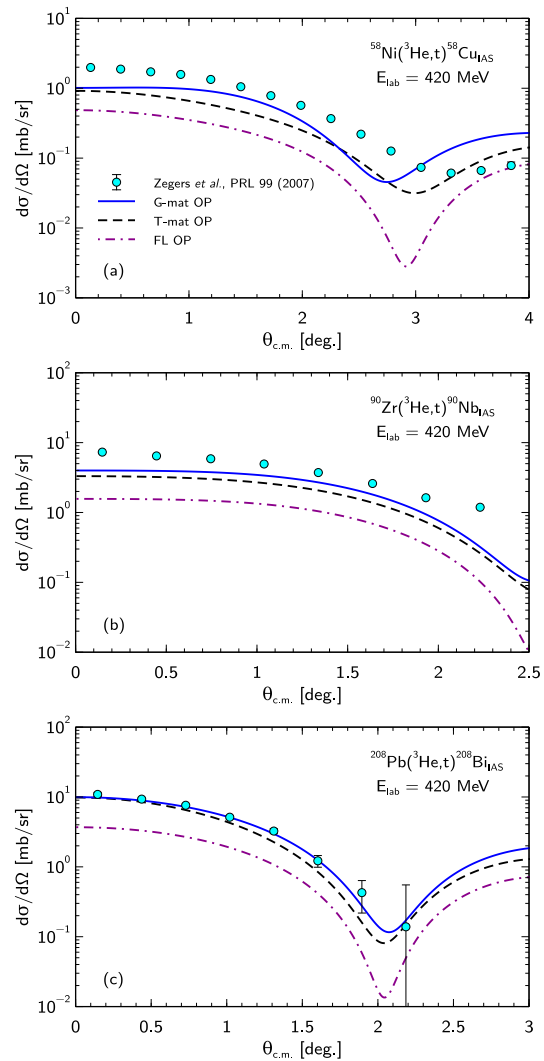}
\caption{Differential cross section of the charge-exchange ($^3$He,$t$)IAS reactions at forward angles on $^{58}$Ni  (panel a), $^{90}$Zr (panel b) and $^{208}$Pb (panel c) at an incident energy of the $^3$He projectile of 420 MeV in the laboratory frame. Results  are shown for calculations done with the $G$-mat OP,  $T$-mat OP and FL OP,  and compared with the experimental data of Ref.\ \cite{Zegers07}.
}
\label{fig:fig4}
\end{center}
\vskip -0.5cm
\end{figure}

The  differential cross section of the charge-exchange ($^3$He,$t$)IAS reactions on $^{58}$Ni, $^{90}$Zr and $^{208}$Pb at forward angles are shown, respectively, in panels a, b and c of Fig.\ \ref{fig:fig4}, for an incident energy of the $^3$He projectile of 420 MeV in the laboratory frame. Results are shown for calculations performed with the $G$-mat, $T$-mat and FL $^3$He-nucleus OP and compared with the experimental data of Ref.\ \cite{Zegers07}. As in the case of the  ($p,n$)IAS reactions, the best description of the data is given by the $G$-mat OP, proving once more the important role played by the  density dependence of the effective NN interaction.  Note that the calculation performed with the $T$-mat OP provides also a reasonable description of the experimental cross sections whereas the FL OP underestimates them for the three reactions considered.  A very good agreement between the data and the prediction of the  $G$-mat OP is obtained in particular for the case of the $^{208}$Pb target. Note also that, in this case, the $G$-mat OP and the $T$-mat OP predict very similar differential cross sections for scattering angles $\leq 0.5\degree$. This is an indication that,  at very forward angles, the differential cross section of the $^{208}$Pb($^3$He,$t$)$^{208}$Bi$_{\rm IAS}$ reaction is not very sensitive to the density dependence of the effective NN interaction.  This is not the case for the other two reactions where there is a clear difference between the $G$-mat OP and the $T$-mat OP results in the whole range of angles considered.

We would like to point out here that the reactions $^{90}$Zr($^3$He,$t$)$^{90}$Nb$_{\rm IAS}$ and $^{208}$Pb($^3$He,$t$)$^{208}$Bi$_{\rm IAS}$ can be used to deduced the neutron skin thickness of $^{90}$Zr ($\Delta R_{np}$($^{90}$Zr)) and $^{208}$Pb ($\Delta R_{np}$( $^{208}$Pb)). In Ref.\ \cite{Loc14} for instance, the experimental differential cross sections of these two reactions were used to adjust the radial parameter of a empirical neutron density \cite{Ray78} to obtain the best DWBA fit to the data, finding, respectively, the values $\Delta R_{np}$($^{90}$Zr$)=0.09\pm 0.03$ fm and $\Delta R_{np}$( $^{208}$Pb$)=0.16\pm0.04$ fm, compatible with nuclear structure calculations and current existing data \cite{Grasso01,Yako06,PREX12}. In our calculation, as mentioned at the end of Sec.\ \ref{sec:formalism}, we use densities for the target nuclei obtained from the SHF calculations  of Ref.\ \cite{Goriely07}.  These SHF calculations predict $\Delta R_{np}$($^{90}$Zr$)=0.07$ fm and $\Delta R_{np}$( $^{208}$Pb$)=0.16$ fm, respectively, in agreement with the results of Ref.\ \cite{Loc14}. It is notable that in our calculation no free parameter have to be adjusted to get  a good description of the differential cross sections and,  simultaneously, reasonable values for the neutron skin thickness of $^{90}$Zr and $^{208}$Pb.

\begin{figure}[t!]
\begin{center}
\includegraphics[width=7.0cm,angle=0,clip]{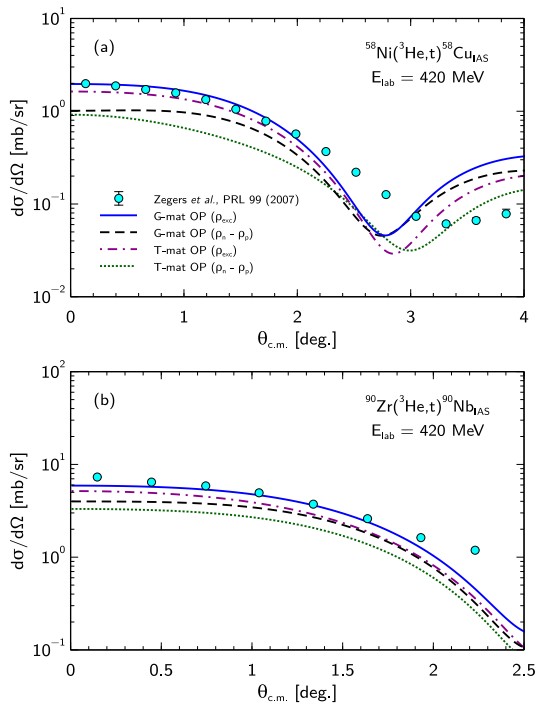}
\caption{Differential cross sections of the $^{58}$Ni($^3$He,$t$)$^{58}$Cu$_{\rm IAS}$ (panel a) and $^{90}$Zr($^3$He,$t$)$^{90}$Nb$_{\rm IAS}$ (panel b) charge-exchange reactions at 
$E_{lab} = 420$ MeV obtained using the neutron excess density in the calculation of the transition densities.  
Results using the difference between neutron and proton densities are also shown
for comparison. Calculations have been done using the $G$-mat and $T$-mat OPs.
Experimental data is taken from Ref.\ \cite{Zegers07}.
}
\label{fig:fig5}
\end{center}
\vskip -0.5cm
\end{figure}

It was shown by Auerbach and Van Giai \cite{Auerbach81}  that if the excess neutron number ($N-Z$) in a nucleus is small then the determination of the transition densities in CXIAS should be done using the neutron excess density ($\rho_{exc}$) instead of the difference between the neutron and proton densities ($\rho_n-\rho_p$). Following these authors we have calculated again the differential cross section of the $^{58}$Ni($^3$He,$t$)$^{58}$Cu$_{\rm IAS}$ and  $^{90}$Zr($^3$He,$t$)$^{90}$Nb$_{\rm IAS}$ reactions using $\rho_{exc}$ in the determination of the transition densities. In practice, we have simply replaced in Eq.\ (\ref{folcalIV}) the isovector density matrix of the target nucleus by the corresponding $\rho_{exc}$ while keeping  the difference $\rho_n-\rho_p$ for the $^3$He projectile. The results, shown in Fig.\ \ref{fig:fig5}, are compared with those obtained previously with the $G$-mat OP using the difference $\rho_n-\rho_p$, reported already in Fig.\ \ref{fig:fig4}, and with those obtained with the $T$-mat OP using both $\rho_{exc}$ and  $\rho_n-\rho_p$. As one can see, when the $G$-mat OP and the $\rho_{exc}$ are used, a better description of the experimental data is achieved in the case of the $^{58}$Ni($^3$He,$t$)$^{58}$Cu$_{\rm IAS}$ reaction, in particular for angles $\leq 2\degree$. However, the data for larger angles is still poorly described. Note that, although the excess neutron number is $10$ in the case of $^{90}$Zr, a remarkable improvement of the description of the differential cross section of the $^{90}$Zr($^3$He,$t$)$^{90}$Nb$_{\rm IAS}$  reaction is also attained.

\section{Summary and Conclusions}
\label{sec:conclusions}

Using the DWBA together with the $G$-matrix double-folding method, employed to determine the nucleus-nucleus optical potential within the framework of the Lane model, in this work we have calculated the differential cross sections of ($p,n$) and ($^3$He,$t$) charge-exchange reactions leading to the excitation of the isobaric analog state  of the target nucleus. $G$-matrices have been obtained from a Brueckner--Hartree--Fock calculation using the Argonne Av18 nucleon-nucleon potential. A good description of the experimental data of the  ($p,n$)IAS and  ($^3$He,$t$)IAS reactions on several targets at different incident energies of the proton and $^3$He projectiles has been found without adjusting any free parameter in our model.  Particularly, a very good agreement between the results obtained with the $G$-mat OP and the experimental data has been obtained in the case of the $^{208}$Pb($^3$He,$t$)$^{208}$Bi$_{\rm IAS}$ reaction.  A clear improvement in the description of the differential cross sections for the $^{58}$Ni($^3$He,$t$)$^{58}$Cu$_{\rm IAS}$ and  $^{90}$Zr($^3$He,$t$)$^{90}$Nb$_{\rm IAS}$ reactions has been found when the neutron excess density is used to determine the transition densities. Our results have shown, in general, that the density and isospin dependences of the nuclear $G$-matrices play a non-negligible role in the description of the experimental data.

Current data on CXIAS reactions comes mainly from measurements on stable target nuclei. Therefore, future measurements of IAS transitions on neutron-rich nuclei, such as those that will be carried out at  the Facility for Rare Isotope Beams (FRIB) at Michigan State University (MSU) are very much awaited for. High-quality data from these measurements will provide very valuable information that will allow us to improve our presenty knowledge on the nuclear force and, in particular, of its isospin dependence. 
 
\section*{Acknowledgements}

B. M. L. would like to thank Prof. Dao T. Khoa and Prof. K. Amos for the discussions and encouragement at the initial state of the work.
This work has been partially supported by the Vietnam National Foundation for Science and Technology Development (NAFOSTED) and by the COST Action CA16214.


\end{document}